\documentclass[reprint, showpacs, showkeys, superscriptaddress, aps, pre, twocolumn, 10pt]{revtex4-1}

\usepackage[T1]{fontenc}
\usepackage[utf8]{inputenc}
\usepackage{amsmath}
\usepackage{amssymb}
\usepackage{enumitem}
\usepackage{graphicx}
\usepackage{refstyle}
\usepackage{multirow}
\usepackage[normalem]{ulem}
\newcommand\redout{\bgroup\markoverwith
{\textcolor{red}{\rule[.5ex]{5pt}{0.5pt}}}\ULon}
\usepackage[
  citecolor=blue,
  colorlinks,
  linkcolor=blue,
  urlcolor=blue,
]{hyperref}
\usepackage[dvipsnames]{xcolor}

\newcommand{\be}{\begin{equation}}
\newcommand{\ee}{\end{equation}} 
\newcommand{\bea}{\begin{eqnarray}}
\newcommand{\eea}{\end{eqnarray}}

\begin{document}
\title{Weight of fitness deviation governs strict physical chaos in replicator dynamics}
\author{Varun Pandit}
\email{varpan.varun@gmail.com}
\affiliation{
  Department of Physics,
  Indian Institute of Technology Kanpur,
 Uttar Pradesh 208016, India
}
\author{Archan Mukhopadhyay}
\email{archan@iitk.ac.in}
\affiliation{
 Department of Physics,
  Indian Institute of Technology Kanpur,
  Uttar Pradesh 208016, India
}
\author{Sagar Chakraborty}
\email{sagarc@iitk.ac.in}
\affiliation{
  Department of Physics,
  Indian Institute of Technology Kanpur,
  Uttar Pradesh 208016, India
}

%
%
%
%
%
%
%
%
%

\begin{abstract}
Replicator equation---a paradigm equation in evolutionary game dynamics---mathematizes the frequency dependent selection of competing strategies vying to enhance their fitness (quantified by the average payoffs) with respect to the average fitnesses of the evolving population under consideration. In this paper, we deal with two discrete versions of the replicator equation employed to study evolution in a population where any two players' interaction is modelled by a two-strategy symmetric normal-form game. There are twelve distinct classes of such games, each typified by a particular ordinal relationship among the elements of the corresponding payoff matrix. Here we find the sufficient conditions for the existence of asymptotic solutions of the replicator equations such that the solutions---fixed points, periodic orbits, and chaotic trajectories---are all strictly physical, meaning that the frequency of any strategy lies inside the closed interval zero to one at all times. Thus, we elaborate which of the twelve types of games are capable of showing meaningful physical solutions and for which of the two types of replicator equation. Subsequently, we introduce the concept of the weight of fitness deviation that is the scaling factor in a positive affine transformation connecting two payoff matrices such that the corresponding one-shot games have exactly same Nash equilibria and evolutionary stable states. The weight also quantifies how much the excess of fitness of a strategy over the average fitness of the population affects the per capita change in the frequency of the strategy. Intriguingly, the weight's variation is capable of making the Nash equilibria and the evolutionary stable states useless by introducing strict physical chaos in the replicator dynamics based on the normal-form game.
\end{abstract}
\maketitle
\section{Introduction}
In spite of more realistic statistical frameworks that include effects of finite-size effects~\cite{pastor2002epidemic} and  restricted interactions among individuals in a population~\cite{traulsen2007pairwise}, collection of deterministic formulations~\cite{nowak2004evolutionary} of evolutionary dynamics continues to be a fruitful and important approach in understanding the evolutionary processes mathematically. In fact, some such very different formulations are part of a single unified framework~\cite{page2002unifying}. Replicator equation~\cite{taylor1978evolutionary,cressman2014replicator} is one such formulation of evolutionary game theory---a phenotypic approach to evolutionary dynamics. For an evolving population of many types (read phenotypes or strategies), the replicator equation provides cause-effect relationship between the frequency (relative abundance) of a type and the fitness (measure of reproductive success) of the type: the per capita change in frequency of a particular type should be positive/negative when its fitness is more/less than the average fitness of the population as a whole. In absence of any mutation process, as is the case with replicator equation, constant fitness functions for the types doesn't give any nontrivial dynamics. However, even with the simplest type of complication, viz., fitnesses are linear functions of the frequencies, replicator equation becomes nonlinear and consequently exhibits dynamically rich behaviour including chaos~\cite{schnabl1991fullchaos,skyrms1992chaos, chawanya1996chaos,sato2002chaos,sato2003coupled,taekho2017pre}.

The replicator equation is very straightforward and intuitive---all it does is to model, in the simplest possible way, selection of the fitter traits followed by their replication (reproduction), hence the name. Thus, it is not surprising that it finds applications in other processes analogous to evolution such as autocatalytic reactions in networks~\cite{stadler2003molecular} and reinforcement learning~\cite{Borgers1997learning} apart from in economics~\cite{silverberg1997evolutionary} and social systems~\cite{wang2009emergence,helbing2010evolutionary}. Evolutionary game theory has been successful in explaining the predominance of certain traits and (biological or social) behaviours of many living organisms. In any organism, including human beings, assumptions of strong rationality almost always fails. However, under the more realistic setting where one should consider the players of a game only boundedly rational, their survival hinges on their ability to learn optimal strategies as the game is repeatedly played. In fact, such a learning process for the case of simple two player zero-sum rock-paper-scissors game leads to low dimensional Hamiltonian chaos when modelled using replicator equation~\cite{sato2002chaos}. Appearance of chaos in evolutionary game theory means that Nash equilibria and evolutionary stable strategies are not decisive in determining the final fate of the corresponding games.

Of course, that the replicator equation by virtue of its inherent nonlinear nature shows chaos is not surprising at all from a mathematical point of view. What we find interesting in this paper is that certain discrete (in time) version of the replicator equation is capable of showing chaos in two-player, two-strategy games, something impossible in continuous replicator equations because of the restriction imposed by Poincar\'e--Bendixson theorem~\cite{poincare1881courbes,bendixson1901courbes}. Moreover more than one discrete versions of  the replicator equation with drastically different dynamical behaviours exist. \textcolor{black}{It may be remarked that even in a two-player, two-strategy discrete monotone selection dynamics, which is a superset of replicator dynamics, chaos may be observed~\cite{cressman2003evolutionary}.} Additionally, as we show see in this paper, the discrete replicator equations are more subtle because they lead to unphysical solutions for the frequencies meaning that frequency values lie outside the closed interval zero to one. Consequently, one is restricted to use only such kind of fitness functions so that the solutions are always physical making a priori identification of such functions indispensable. This is what we achieve in this paper. A related earlier work~\cite{vilone2011chaos} didn't contrast dynamics of different discrete replicator equations nor did that go much into the important detailed analytical investigation of the condition for the existence of realistic physical solutions. After discussing this issue of physical solutions in Sec.~\ref{sec:3}, we show in Sec.~\ref{sec:3.B} how chaos can appear only selectively in evolutionary game dynamics depending on what kind of discrete replicator one is working with. 

An even more interesting result follows in Sec.~\ref{sec:5} where we see that the difference of a type's fitness from the average fitness, when magnified by a factor (which we quantify by the \emph{weight of fitness deviation} to be discussed in the paper), not only effects the rate of per capita change of a trait frequency but can also render Nash equilibria and evolutionary stable strategies or states useless by introducing physical chaotic solutions. In fact, it is the weight of fitness deviation that is seen to govern the strict physical solutions, including chaotic solutions, in evolutionary game dynamics with a given payoff matrix for the corresponding replicator game dynamics. This is another important result of our present work. 

However, before embarking on the technical discussion of the results, in the very next section we elaborate on some of the relevant game theoretic concepts that are going to be extensively used later in this paper. 

\section{Relevant game theoretic concepts and replicator equations}
Non-cooperative game theory studies the strategic interaction of independent individuals with no enforcement of cooperation due to an external agent~\cite{osborne2004introduction}. A non-cooperative game is represented by the set of $N$ players, each with its own set of strategies and a utility function. For each individual, the utility function is the map from the set of strategy profiles to real line $\mathbb{R}$. Recall that a strategy profile (also called strategy combination) is an $N$-tuple which includes one strategy of each player. The utilities of symmetric normal form game is represented by a payoff matrix. Let $\textsf{U}$ be the payoff matrix of $N$ strategy symmetric game,  ${\mathcal{S}^{N}}$ be the pure strategy set and the normalized (non-negative) frequency of strategy  $s_{i} \in {\mathcal{S}^{N}}$ be given by $p_{i}$. Let the $N$-simplex $\Sigma^{N}$ be the set of all mixed strategies which by definition includes the pure strategies. It is customary to call the strategies for which $p_i>0\,\forall i\in\{1,2,\cdots,N\}$, completely mixed strategies. Symbolically, $\textsf{U}({\textbf{p},\textbf{q}})$ gives utility for a player playing the mixed strategy $\textbf{p}$ against the mixed strategy $\textbf{q}$. 

Having set up the ideas behind a game, let us introduce two very well-known and extremely useful concepts: (i) a mixed strategy $\textbf{p} \in \Sigma^{N}$  is a \emph{Nash equilibrium} if $\textsf{U}(\textbf{p},\textbf{p}) \ge \textsf{U}(\textbf{q}, \textbf{p})$ $\forall\,\textbf{q} \in \Sigma^{N}$; and (ii) a mixed strategy $\textbf{p} \in \Sigma^{N}$ is an \emph{evolutionary stable strategy} (ESS)~\cite{smith1982evolution} if $\forall\,\textbf{q} \in \Sigma^{N}$, either 
$\textsf{U}(\textbf{p},\textbf{p}) > \textsf{U}(\textbf{q},\textbf{p})$ or  $\textsf{U}(\textbf{p},\textbf{q}) > \textsf{U}(\textbf{q},\textbf{q})$ when $\textsf{U}(\textbf{p},\textbf{p}) = \textsf{U}(\textbf{q},\textbf{p})$.
In other words, {Nash equilibrium} corresponds to the strategy profile (which now includes completely mixed strategies) from which no player has any incentive to unilaterally deviate and  an ESS is a strategy that cannot be invaded by any initially rare alternative strategy. It may be emphasized that the word `rare' is in context of a population of players playing a particular strategy.

 We are now well-equipped to introduce the definitions of \emph{ordinally equivalent games} and \emph{cardinally equivalent games} that have been extensively used in this paper to understand the dynamics of games of interest.
 \begin{itemize}
\item Two $N$-strategy symmetric normal form games with payoff matrices $\textsf{U}^{(1)}$ and $\textsf{U}^{(2)}$ are said to be ordinally equivalent games if both games consist of same players with same set of pure strategies, $\mathcal{S}^N$, such that $\textsf{U}^{(1)}(s_{i},s_{j})\ge\textsf{U}^{(1)}(s_{i'},s_{j'})$ implies (and is implied by) $\textsf{U}^{(2)}(s_{i},s_{j})\ge\textsf{U}^{(2)}(s_{i'},s_{j'})$ for all $s_{i},s_{j},s_{i'},s_{j'}\in \mathcal{S}^N$. The payoff matrices of ordinally equivalent games are ordinally equivalent utilities and one payoff matrix is strictly increasing transformation of the  other~\cite{hammond2005utility}.

\item Two $N$-strategy symmetric normal form games with payoff matrices $\textsf{U}^{(1)}$ and $\textsf{U}^{(2)}$ are said to be cardinally equivalent games if both games consist of same players with same set of mixed strategies, $\Sigma^N$, such that $\textsf{U}^{(1)}(\textbf{p}_{i},\textbf{p}_{j})\ge\textsf{U}^{(1)}(\textbf{p}_{i'},\textbf{p}_{j'})$ implies (and is implied by) $\textsf{U}^{(2)}(\textbf{p}_{i},\textbf{p}_{j})\ge\textsf{U}^{(2)} (\textbf{p}_{i'},\textbf{p}_{j'})$ for all $\textbf{p}_{i},\textbf{p}_{j},\textbf{p}_{i'},\textbf{p}_{j'}\in \Sigma^N$. Pay-off matrices of  cardinally equivalent games are \emph{positive} affine transformation of each other. {It should be noted that cardinally equivalent games have same set of Nash equilibria and evolutionary stable strategies.}
\end{itemize}
 It is clear that two cardinally equivalent games are ordinally equivalent as well but the converse is generally not true.

Coming to the main topic of this paper, we are interested in the evolution of the frequencies of strategies in a population of players for whom natural selection is the only mechanism driving the changes in the frequencies. H. Spencers' famous phrase `survival of the fittest' hints at \emph{differential} reproduction based on \emph{fitnesses} of the players. Qualitatively speaking, fitness of an individual is measured by the individual's ability to survive and subsequently reproduce, thereby passing the trait to progeny~\cite{orr2009fitness} whose trait would, thus, be that of the parent. Under natural selection a trait is selected if it outperforms the average fitness of the population. Change in the frequency of a trait type in the strategic interaction of a population is modeled using a replicator equation. In what follows, we present how researchers have mathematized these ideas.

Let a population consist of $n$ types corresponding to $n$ points (${\textbf{p}_{1},\textbf{p}_{2},\cdots,\textbf{p}_{n}}$)  on $\Sigma^{N}$~\cite{hofbauer1998evolutionary}.  
Let ${\sf{\Pi}}$ corresponds to the fitness matrix of a population with $n$ interacting types. The payoff of the $i^{\rm th}$ type against the $j^{\rm th}$ type is given by ${{\pi}}_{ij}=\textbf{p}_{i}^T\textsf{U}\textbf{p}_{j}$---the element of $\sf{\Pi}$ located at $i^{\rm th}$ row and $j^{\rm th}$ column. The state of a population, $\textbf{x}$, is defined by the set of normalized frequencies $\{ x_{1},...,x_{n} \}$ on $n$-simplex $\Sigma^{n}$ for each type in the population. It may be noted that the {average population strategy} $\bar{\mathbf{p}}=\sum_ix_i\textbf{p}_i$ traverses a unique orbit in $\Sigma^N$ depending on the evolution of $\mathbf{x}$ in $\Sigma^n$. A replicator equation, discussed below, is designed to model the change in frequency of a state--- it increases the frequency of a better performing type (trait) and decreases the frequency of  under-performing type (trait)~\cite{taylor1978evolutionary}. The performance is measured relative to the average fitness of population. The expected fitness of $i^{th}$ type is $\left({\sf{\Pi}} \textbf{x}\right)_{i}$ and the average fitness of population is $\textbf{x}^T {\sf{\Pi}} \textbf{x}$. Replicator equation increases the frequency of $i^{th}$ type if its relative fitness is positive. 

Let $\textbf{f}(\textbf{x}):\Sigma^n\rightarrow\Sigma^n$ be the discrete model of replicator dynamics with components $f_{i}$ $(i=1,2,\cdots, n)$ . Two of the common forms~\cite{Weissing1991evolutionary, dekel1992optimazing, cabrales1992selection, nowak1993chaos, vilone2011chaos} of the discrete replicator maps in vouge in the research literature are:
\begin{eqnarray}
&&x'_{i}={f}_{i}(\textbf{x})=x_{i}+ x_{i}\left[ \left( {\sf{\Pi}} \textbf{x} \right)_{i}-\textbf{x}^{T}{\sf{\Pi}} \textbf{x} \right],
\label{eq:type-I}\\
&&x'_{i}={f}_{i}(\textbf{x})=x_{i}\frac{({\sf{\Pi}} \textbf{x})_{i} }{\textbf{x}^{T}{\sf{\Pi}} \textbf{x}}\,,\label{eq:type-II}
\end{eqnarray}
where $i$ can run from $1$ to $n$. $x'_i$ essentially means the value $x_{i}$ at the very next instant of time being sampled discretely. For convenience, we term Eq.~\ref{eq:type-I} as `type-I' replicator equation and Eq.~\ref{eq:type-II} as `type-II' replicator equation.  

A state $\textbf{x}$ is an \emph{evolutionary stable state} of population if there exists a neighbourhood, $\mathcal{B}_\mathbf{x}$, of $\textbf{x}$ such that $\forall\,\textbf{y} \in \mathcal{B}_\mathbf{x}\backslash\{\mathbf{x}\}$, state $\textbf{x}$ in not invaded by $\textbf{y}$~\cite{hofbauer1998evolutionary}, i.e.,
\begin{equation}
\textbf{x}^{T} {\sf{\Pi}} \textbf{y}>\textbf{y}^{T}{\sf{\Pi}}\textbf{y}\hspace{0.5cm} \forall \,\textbf{y} \in \mathcal{B}_\mathbf{x}\backslash\{\mathbf{x}\}\,.
\end{equation}
Under natural selection, an evolving population characterised by the underlying strategy profile (${\textbf{p}_{1},\textbf{p}_{2},\cdots,\textbf{p}_{N}}$) is expected to reach a robust composition specified by $\mathbf{x}$ that is evolutionary stable. Though there is the possibility that it may never happen.
Note that just in line with the definition of Nash equilibrium introduced earlier,  $\mathbf{x}$ is called a \emph{Nash equilibrium} (NE) if $\textbf{x}^{T} {\sf{\Pi}} \textbf{x}\ge\textbf{y}^{T}{\sf{\Pi}}\textbf{x}$ $\forall\,\mathbf{y}\in\Sigma^n$. This NE is, by construction, symmetric.
\section{Strict Physical Regions}
\label{sec:3}
\begin{figure}
\includegraphics[scale=0.26]{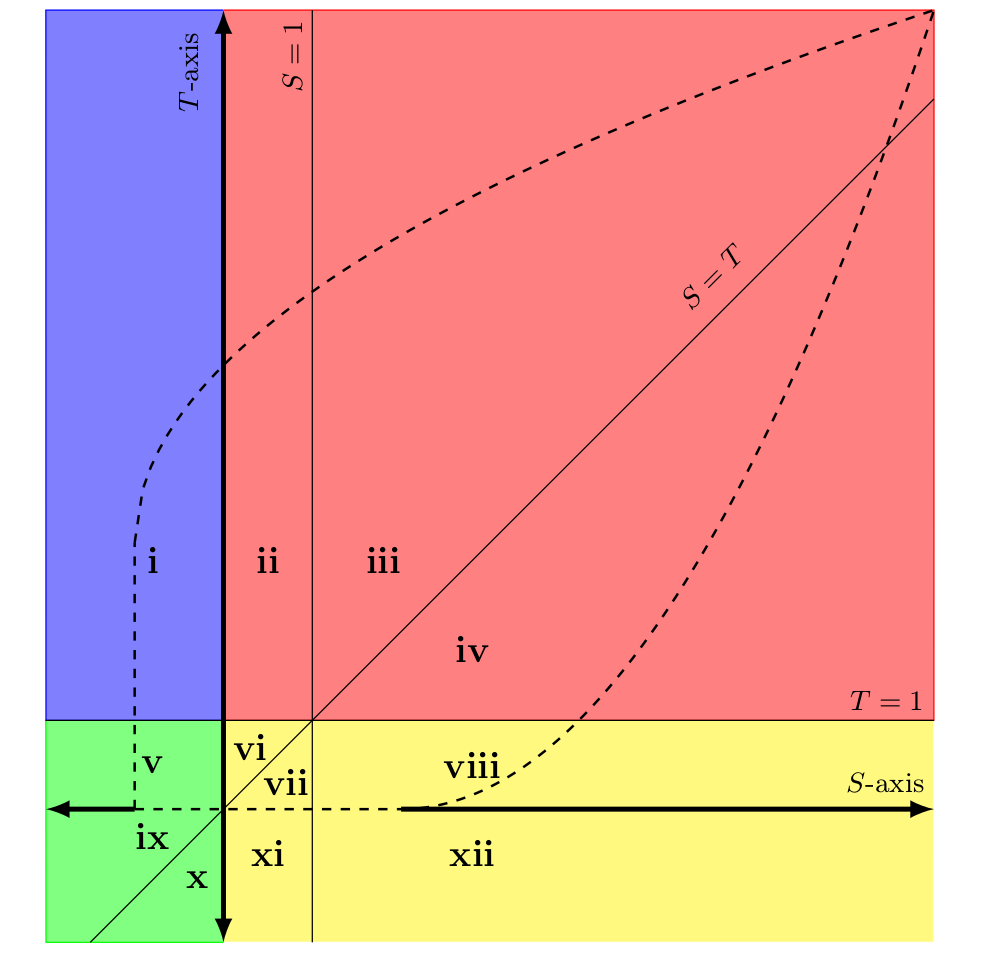}
\caption{Classification of twelve classes of ordinally equivalent games. The straight lines $S=0,\,T=0,\,S=1,\,T=1,$ and $S=T$ separates $S$-$T$ space into twelve non-overlapping regions each consisting of a set of ordinally equivalent games. \textcolor{black}{These twelve sets correspond to the well-known games, viz., games of Prisoner's Dilemma (\textbf{i}), Chicken games  (\textbf{ii}), Leader games (\textbf{iii}), games of Battle of Sexes (\textbf{iv}), Stag-hunt games (\textbf{v}), Harmony I games (\textbf{vi}), Harmony II games (\textbf{vii}),Deadlock II games (\textbf{viii}), Coordination I games (\textbf{ix}), Coordination II games (\textbf{x}), Harmony III games (\textbf{xi}), Deadlock I games (\textbf{xii}).} When played as one-shot game, the games in red region, blue region, and yellow region consist of only one symmetric NE whereas the games in green region consist of three symmetric Nash equilibria. The dashed closed curve is the curve inside (including boundaries) which type-I replicator map gives strict physical solutions. Type-II replicator map's strict physical region is given for the region where both $S$ and $T$ are non-negative.}
\label{fig_1}
\end{figure}

From now on we shall focus on the aforementioned two types of discrete replicator models (Eq.~(\ref{eq:type-I}) and Eq.~(\ref{eq:type-II})) and their dynamics on $2$-simplex $\Sigma^{2}$. The twelve ordinally equivalent classes of two strategy symmetric normal form games can be represented~\cite{hauert2001effects,hummert2014molbiostat} by the following generalized payoff matrix: 
\begin{eqnarray}
\begin{tabular}{c}
 ${\sf{\Pi}}=\textsf{A}$=
 $\begin{bmatrix}  
1 & S \\ T & 0
\end{bmatrix}$;\quad $S,T\in\mathbb{R}$\,.
\end{tabular}\vspace{2mm}
\label{eqn:PayOff_A}
\end{eqnarray}
In $S$-$T$ plane, these twelve classes are demarcated by the straight lines: $T=0,T=1,T=S,S=0\,,\&\,S=1$ (see regions \textbf{i} to \textbf{xii} in Fig.~\ref{fig_1}).  It should be remarked that a two-strategy game with payoff matrix having equal diagonal elements (e.g., a symmetric coordination game with two Pareto optimal symmetric Nash equilibria) is not represented by $\sf{A}$\textcolor{black}{; in the context of this paper, a short but sufficient discussion on such games has been presented in Appendix~\ref{Ap:1}}.

\textcolor{black}{Only one symmetric NE exists in Prisoner's dilemma (\textbf{i}), Harmony I game (\textbf{vi}), Harmony II game (\textbf{vii}), Deadlock II game (\textbf{viii}), Harmony III game (\textbf{xi}), and Deadlock I game (\textbf{xii})}. This NE is strict and hence is also an ESS. Chicken game (\textbf{ii}), Leader game (\textbf{iii}), and {Battle of sexes} (\textbf{iv}) have only one symmetric NE that happens to be a completely mixed NE and ESS.  The rest three games Stag-hunt (\textbf{v}), \textbf{ix}, and \textbf{x} have three symmetric Nash equilibria --- one completely mixed and two strict. Only the strict ones are ESS's. 

Before we proceed further, let us note a subtle point. Mathematically speaking, for arbitrary ${\sf \Pi}$, both type-I and type-II equations can give a mathematical solution that traces a trajectory in the $n$-dimensional phase space $\mathbb{R}^n$ but is not bounded in $\Sigma^n\subset \mathbb{R}^n$. This is physically not allowed because normalised $x_i$, by definition, must remain in the interval $[0,1]\,\forall i\in\{1,2,\cdots,n\}$ at all times; otherwise such solutions are termed as {unphysical} solutions. We are mostly interested in physical solutions but the above-mentioned subtlety motivates us to introduce the following concepts related to the physical solutions (solutions that are not unphysical): We say a replicator equation $\textbf{f}$ has \emph{strict physical solutions} if  $\textbf{f}:\Sigma^{n} \rightarrow \Sigma^{n}$ for all initial $\textbf{x}\in \Sigma^{n}$ at all times. Since the dynamics of the replicator maps must depend on parameters $S$ and $T$, we now introduce the concepts of {strict} physical region in $S$-$T$ space: A region---subset of $\mathbb{R}^2$---in $S$-$T$ space is \emph{strict physical region} for a particular replicator equation if the equation has strict physical solutions for all points $(S,T)$ in that region for every initial conditions. This concept is of central-most importance in this paper.

In what follows we now rigorously find the strict physical regions for both type-I and type-II equations written using payoff matrix ${\sf A}$:
\subsection{Type-I replicator equation}
In two strategy game the state of population is $\textbf{x}=\{x_{1},x_{2}\}$.  Let $x_{1}=x$ and $x_{2}=1-x$. The components of type-I replicator map for payoff matrix ${\sf{A}}$ are therefore given by:
\begin{equation}
f_{1}(\textbf{x})=x_{1}+ x_{1}\left( \left( {\sf{A}} \textbf{x} \right)_{1}-\textbf{x}^{T}{\sf{A}} \textbf{x} \right) \,; f_{2}(\textbf{x})=1-f_{1}(\textbf{x})\,.
\label{eq:type-IA}
\end{equation}
Let the function $f_{1}(x_{1},x_{2})$ of type-I replicator equation in Eq.~(\ref{eq:type-IA}) be called $f_{I}(x)$.  Expanded form of $f_{I}(x)$ as a polynomial is given as:
\begin{equation}
f_{I}(x)=(S+T-1)x^3+(1-2S-T)x^2+(1+S)x\,,
\label{eq:type-IAExpand}
\end{equation}
which possibly can have one maximum and one minimum.
Map $x'=f_{I}(x)$ has at most three fixed point solutions: $x=0$, $x=1$, and $x=x_{(m)}={S}/{(S+T-1)}$. 

Now to find the strict physical region for the equation $x'=f_I(x)$, we note the following facts:
\begin{enumerate}
\item Whenever $df_{I}(0)/dx=1+S<0$, type-I equation doesn't give strict physical solution in the region $S<-1$ because then the replicator equation maps all points in some small neighbourhood of $x=0$ outside the simplex such that $f_I(x)<0$. 
\item Again, if $df_{I}(1)/dx=T<0$, then type-I equation doesn't give strict physical solution because each point $x$ in some small neighbourhood of $x=1$ is mapped outside the simplex such that $f(x)>1$.
\item If $1+S\ge0$ (cf. point 1 above) and $T\ge0$ (cf. point 2 above), and additionally there exists a point of inflection $x^{*}=-\frac{(1-2S-T)}{3(S+T-1)}$ (i.e., $d^2f_{I}(x^*)/dx^2=0$) of cubic polynomial $f_{I}$ such that it lies outside the simplex i.e. $x^{*} \notin [0,1]$, then the map is monotonically non-decreasing on simplex and type-I gives strict physical solution. (However, when either $1+S=0$ or $T=0$, one should also ensure that $d^2f_{I}(0)/dx^2\ge0$ and $d^2f_{I}(1)/dx^2\le0$ respectively for the existence of strict physical regions.)
\item However, if $1+S\ge0$ and $T\ge0$ (cf. point 1 and 2 above), and point of inflection $x^{*}$ lies on the simplex, then there are two possible cases:
\subitem 4a. When $df_{I}(x^{*})/dx= 1+S-\frac{\left(1-2S-T\right)^2}{3\left(S+T-1\right)}\ge0$, no points of maximum or minimum exist inside the simplex $\Sigma^2$ and the map has only strict physical solutions. This is so because if the slope at the point of inflexion is nonnegative then the slope is non-negative for all $\{x,1-x\}\in\Sigma^2$ and hence function $f_I$ is monotonically non-decreasing.
\subitem 4b. When $df_{I}(x^{*})/dx<0$, both the points of minimum and maximum, say $x_\textrm{min}$ and $x_\textrm{max}$, lie on the simplex. The map has strict physical solution if  both $f_I(x_\textrm{min})$ and $f_I(x_\textrm{max})$ lie on the simplex which is mathematically implied by $(S-2)^2-4T\le0$ and $(T-3)^2-4(1+S)\le0$, respectively. These inequalities are arrived at by demanding that both cubic equations $f_{I}(x)=0$ and $f_{I}(x)=1$ should not have three distinct real solutions---otherwise, $f_I(x_\textrm{min})$ or $f_I(x_\textrm{max})$ respectively lie outside the simplex.
\end{enumerate}
All the inequalities as discussed above straightforwardly yield a leaflike region (see Fig.~\ref{fig_1}) in parameter space $S$-$T$ for which type-I replicator equation gives strict physical solutions. 
\subsection{Type-II replicator equation}
Type-II replicator equation for payoff matrix ${\sf{A}}$ can be written as:
\begin{eqnarray}
f_{1}(\textbf{x})=x_{1}\frac{(A\textbf{x})_{1}}{\textbf{x}^T{\sf{A}}\textbf{x}}\,; f_{2}(\textbf{x})=1-f_{1}(\textbf{x})\,.
\label{eq:type-IIA}
\end{eqnarray}
Similar to what has been done with type-I replicator equation, let $f_{II}(x)$ be function $f_{1}(x_{1},x_{2})$ of Eq.~(\ref{eq:type-IIA}) for type-II replicator equation, i.e.,
\begin{equation}
f_{II}(x)=x\frac{x(1-S)+S}{x^2+(S+T)x(1-x)}\,.
\label{eq:type-IIExpand}
\end{equation}
Note that when $S,T\ge0$, $f_{II}\in[0,1]\,\forall\,x\in[0,1]$ because generally in that case the denominator is equal to the positive numerator plus some positive number. Thus, we conclude that type-II replicator equation gives strict physical solutions if $S$ and $T$ both are non-negative. This means that the region given by $(S\ge0,T\ge0)$ is the strict physical region. 
\subsection{Summary and comparison}
The area of the strict physical region $(S\ge0,T\ge0)$ of type-II replicator equation is far bigger than the one (leaf-like region) for type-I replicator equation. \textcolor{black}{Harmony I games (\textbf{vi})  and Harmony II games (\textbf{vii}) lie in both the strict physical regions. Chicken games (\textbf{ii}), Leader games (\textbf{iii}), games of Battle of sexes (\textbf{iv}), and Deadlock II games (\textbf{viii}) always have strict physical physical solutions when type-II replicator equation is used. However, when the type-I replicator equation is used, these games may or may not be in strict physical region. While the classes of Coordination I games (\textbf{ix}), Coordination II games (\textbf{x}), Harmony III games (\textbf{xi}), Deadlock I games (\textbf{xii})  do not fall in either of the strict physical regions, there are strict physical regions for the games of Prisoner's dilemma (\textbf{i}) and Stag-hunt games (\textbf{v}) but only for type-I equation.} 

\section{Replicator dynamics in the strict physical region}
\label{sec:3.B}
\begin{figure}
\includegraphics[scale=0.4]{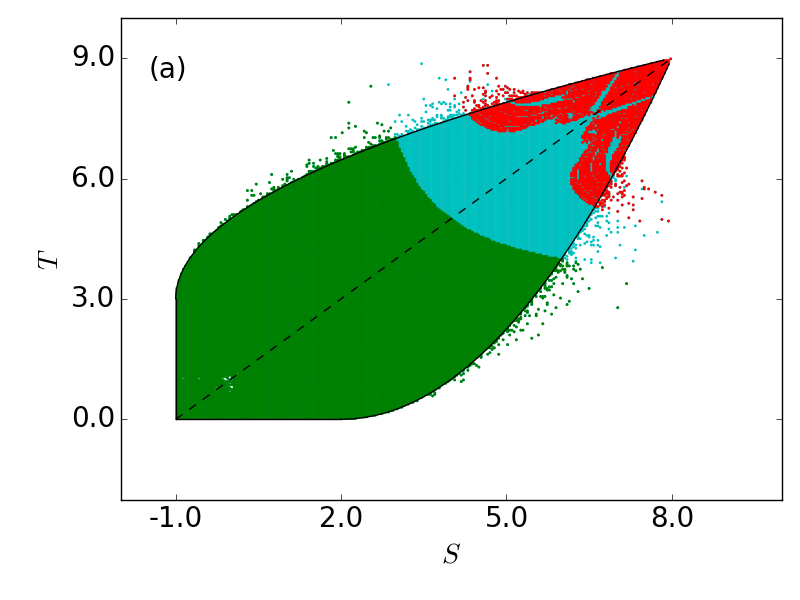} 
\includegraphics[scale=0.4]{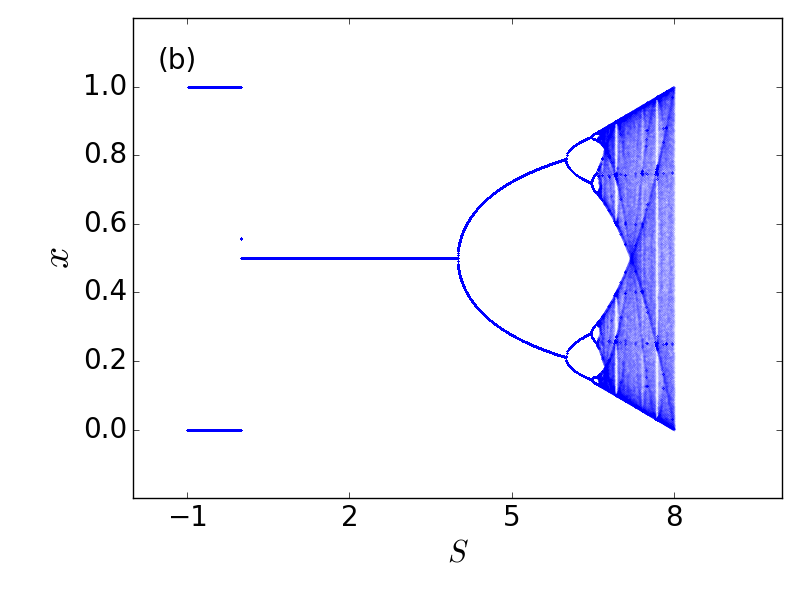}
\caption{Strict physical chaos in replicator dynamics. (a) Attractors [fixed points (green), periodic orbits (cyan) and chaotic orbits (red)] attracting physically strict orbits of type-I replicator equation as concluded by simulating for the Lyapunov exponent for many random initial conditions for every $S,T$ pair on simplex. The sporadic coloured points outside the leaf-like region may either be spurious (a longer numerical run might encounter an unphysical phase point) or be owing to a particular combination of $S$, $T$ and initial $x$. The parameter values in the white region yield unphysical trajectories. (b) Bifurcation diagram showing period doubling route to chaos along $T=1+S$ (dashed black line in subplot~(a)).}
\label{fig_2}
\end{figure}
\begin{table*}
\begin{tabular}{|l|l|l|}
\hline
\textbf{Class of Game} & \textbf{Type-I replicator equation} & \textbf{Type-II replicator equation}\\
\hline
Prisoner's Dilemma ({\bf i}) & Evolutionary Stable State ($x=0$) & --- \\
Chicken game ({\bf ii}) & Evolutionary Stable State ($x = x_{(m)}$) & Evolutionary Stable State ($x=x_{(m)}$)\\  
Leader game ({\bf iii})  &Evolutionary Stable State($x=x_{(m)}$)/Periodic Orbits/Chaos & Evolutionary Stable State ($x=x_{(m)}$) \\ 
Battle of Sexes ({\bf iv}) & Evolutionary Stable State($x=x_{(m)}$)/Periodic Orbits/Chaos & Evolutionary Stable State ($x=x_{(m)}$)\\ 
Stag-hunt game ({\bf v}) & Evolutionary Stable States ($x=0$ or $x=1$) &  --- \\
\textcolor{black}{Harmony I game ({\bf vi})} & Evolutionary Stable State ($x=1$) & Evolutionary Stable State ($x=1$) \\
\textcolor{black}{Harmony II game ({\bf vii})}  & Evolutionary Stable State ($x=1$)& Evolutionary Stable State ($x=1$) \\
\textcolor{black}{Deadlock II game ({\bf viii})}  & Evolutionary Stable State ($x=1$)& Evolutionary Stable State ($x=1$) \\
Games ({\bf ix})-Games ({\bf xii})  & --- & ---\\
\hline
\end{tabular}
\caption{Types of possible attractors of the two discrete replicator equations corresponding to strict physical solutions for the twelve different classes of games. Here, $x_{(m)}={S}/{(S+T-1)}$. An attractor of the replicator equations can be evolutionary stable state (a stable fixed point) or periodic orbit or chaotic.}
\label{tab:1}
\end{table*}
On doing numerical experiments with type-I replicator equation, as shown in Fig.~\ref{fig_2}(a), we observe fixed points, periodic orbits, and chaotic trajectories for the games in the leaf-like strict physical region via period doubling route (Fig~\ref{fig_2}(b)). Physical chaos is observed near the tip of the leaf. Similar numerical experiments with type-II replicator equation doesn't seem to indicate existence of either physical chaotic solutions or periodic orbits. Mostly, we witness that initial conditions asymptotically approaches a fixed point attractor.

Type-I replicator map has been illustrated~\cite{vilone2011chaos} to possess both physical and unphysical chaotic solutions apart from possessing other invariant sets like fixed points and periodic orbits. However, a clear analytical insight into when the solutions are physical or unphysical is unreported in literature to the best of our knowledge. In addition to gathering this insight, we also intend to contrast the dynamics of type-I replicator equation with that of type-II equation. Moreover, this subsection aids in understanding the intriguing results to be discussed in the subsequent subsection.
To this end, for later convenience, henceforth we shall call regions given by $S>0, T>1$; $S<0, T>1$; $S<0, T<1$; and $S>0, T<1$ respectively as quadrant I, quadrant II, quadrant III, and quadrant IV. These quadrants respectively are denoted by the red, the blue, the green, and the yellow regions in Fig.~\ref{fig_1}. 
\subsection{Type-I replicator equation}
For each game in the strictly physical region in quadrants  II and IV, $f_I(x)<x$ and $f_I(x)>x$ $\forall\, x\in(0,1)$ respectively. Therefore, every initial condition (other than the unstable fixed points) converges to attractors $x=0$ and $x=1$ respectively. These fixed point attractors are also evolutionary stable states (see  Appendix~\ref{Ap:2} for a proof). The other possible fixed point $x_{(m)}={S}/{(S+T-1)}$ is unphysical in these quadrants. Also, it is readily seen that any iterate $f^m_I(x)$ ($m\in\mathbb{N}$) of $f_I(x)$ is either less than $x$ (in quadrant II) or greater than $x$ (in quadrant IV) $\forall\, x\in(0,1)$. This implies that in the plot of $x$ vs. $f^m_I(x)$, line $x=f^m_I(x)$ doesn't intersect the graph of $f^m_I(x)$ at any point other than $x=0$ and $x=1$ in the interval $[0,1]$. This means that there are no $m$-period orbits for $m>1$. Moreover, absence of any unstable periodic orbit implies that, by definition, chaotic attractor can not be realized in the state space of the type-I replicator equation with parameter values chosen from the strict physical regions of quadrant II and IV.

$f_I(x)$ for games in strict physical region in quadrant III are monotonically non-decreasing functions for all $x\in[0,1]$. The fixed points $x=0$ and $x=1$ are stable (and thus evolutionary stable state) while the fixed point $x=x_{(m)}$ is physical but unstable. Thus for each initial condition $x\in(0,x_{(m)})$, $f_I(x)<x$ and $f_{I}(x)\in(0,x_{(m)})$, and hence $f^m_I(x)<x,\, \forall m \in \mathbb{N}$. Therefore, no point of periodic orbit exists in interval $(0,x_{(m)})$. Similarly, $f_I(x)>x$ and $f_I(x)\in (x_{(m)},1), \, \forall x \in(x_{(m)},1)$ implying no point of periodic orbit exists in the interval $(x_{(m)},1)$ either. Combining the two results we conclude that type-I replicator equation consists of no periodic orbit --- either stable or unstable---and hence chaotic attractor cannot exist in the strict physical region of quadrant III. 

Unlike in other parts of the strict physical regions, type-I replicator equation for games in strict physical region in quadrant I possess periodic and chaotic orbits. Fixed points $x=0$ and $x=1$ are unstable. Interior fixed point $x_{(m)}$ is physical but shows different stability properties for different games. For games in the strict physical region where $S(T-1)/(S+T-1)<2$~(green region inside leaf-like region in Fig ~\ref{fig_2}(a)), the interior fixed point is locally stable and evolutionary stable state. Type-I equation  undergoes flip bifurcation at $x=x_{(m)}$ and $S(T-1)/(S+T-1)=2$  giving rise to two-period orbit. Subsequently, as $S(T-1)/(S+T-1)$ is continuously varied away from $2$ other higher periodic orbits appear and ultimately chaos~(red region inside leaf-like region in Fig.~\ref{fig_2}(a)) may be arrived at via period-doubling route~(see Fig.~\ref{fig_2}(b)). \textcolor{black}{In passing, we mention that a detailed example of the route to chaos in a cubic map can be found in the 1983 paper of Rogers and Whitley~\cite{rogers1983mathematical}.}
\subsection{Type-II replicator equation}
Only games \textbf{ii}-\textbf{iv} (Chicken games, Leader games, and games of Battle of sexes) and games \textbf{vi}-\textbf{viii} (\textcolor{black}{Harmony I games, Harmony II games, and Deadlock II games}) lie in the strict physical region of the type-II replicator equation. The fixed point $x=1$ exists in all the aforementioned games but the (physical) fixed point $x=x_{(m)}$ exist only for  games \textbf{ii}-\textbf{iv}. $x=x_{(m)}$ and $x=1$ are attractors (and hence evolutionary stable states~\cite{taylor1978evolutionary}) respectively for games \textbf{ii}-\textbf{iv} and games \textbf{vi}-\textbf{viii}.

For all games in the strict physical region, $f_{II}(x)$ is a non-decreasing function inside the simplex. Therefore, for all $x$ in each largest open interval between the consecutive physical fixed points either $f_{II}(x)>x$ or $f_{II}(x)<x$, and $f_{II}(x)$ lies in the same interval. Therefore for each $x$ in each largest open interval either $f^m_{II}(x)>x$ or $f^m_{II}(x)<x,\, \forall m \in \mathbb{N}$. Thus, we conclude that no periodic orbit, and hence no chaotic orbit, is found in the strict physical region of type-II replicator equation.

\subsection{Summary and comparision}
Chaotic and periodic orbits are completely absent in the corresponding strict physical region for type-II equation but they can show up in the strict physical region for type-I replicator equation. Contrary to what has been reported in the literature~\cite{vilone2011chaos}, surprisingly our numerics doesn't show chaos in type-I replicator when ${\sf A}$ corresponds to strict physical region in a Chicken game. However, it could just be that a very particular combination of $S$, $T$, and initial states lead to chaos in chicken game---a possibility not proven to be absent outside the strict physical region.

In Table~\ref{tab:1}, we have summarized the types of attractors possible in both type-I and type-II replicator dynamics for each ordinally equivalent class of games in the strict physical region. 
\section{Cardinally equivalent games and weight of fitness deviation}
\label{sec:5}
\begin{figure}
\includegraphics[scale=0.4]{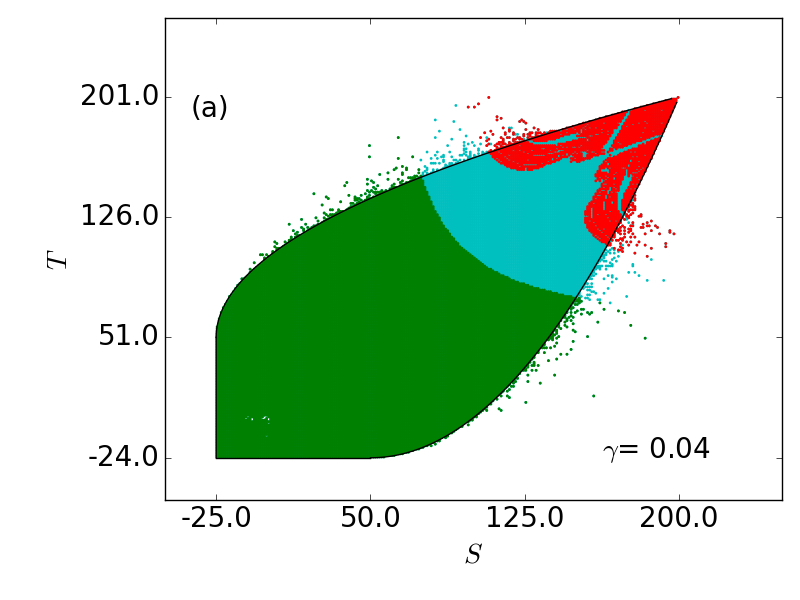}\\
\includegraphics[scale=0.4]{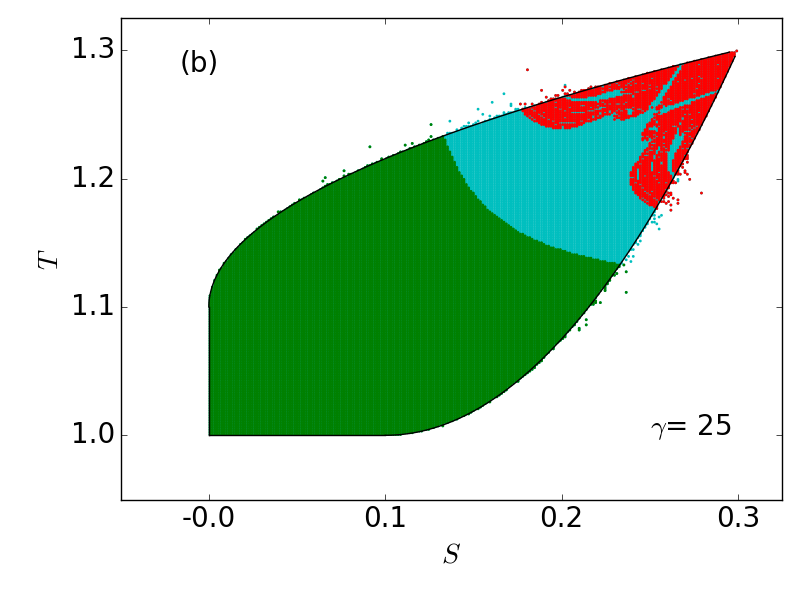} 
\caption{Weight of fitness deviation governs chaos. Types of attractors, viz., fixed points (green), periodic orbits (cyan) and chaotic orbits (red) as obtained by simulating type-I replicator dynamics at (a) $\gamma= 0.04$ and (b) $\gamma=25$. Here as in Fig.~\ref{fig_2}a only those attractors have been marked that attract only physically strict trajectories. Note how $\gamma$ modifies $S$ and $T$ values at which chaos appears.}
\label{fig_3}
\end{figure}
Till now our focus has been on discrete replicator dynamics with payoff matrix ${\sf A}$ that allows for the study of twelve classes of ordinally equivalent games as far as the two-player, two-strategy symmetric games are concerned. In each class there are uncountably infinite number of ordinally equivalent games. {Now, within a class we can shift (by a constant real matix) and scale (by a constant positive number) $\sf{A}$ keeping the characteristics (Nash equilibria, evolutionary stable strategies, etc.) of the one-shot game unchanged.} However, it may be noted that whereas type-I replicator equation is invariant under a shift of the payoff matrix by a constant matrix but \emph{not} under the scale change of the matrix, type-II replicator equation is invariant under the scale change but \emph{not} under the shift. Thus, dynamical behaviours of the solutions to a replicator equation (type-I or type-II) obtained by using two different games that are cardinally equivalent need \emph{not} be identical. 

In view of the above, let us introduce ${\sf \Pi}={\sf{\Pi_2}}$ to be the general $2\times 2$ fitness matrix such that ${\sf{\Pi_2}}$ is a positive affine transformation of payoff matrix $\textsf{A}$, i.e.,
\begin{eqnarray}
\begin{tabular}{c}
${\sf{\Pi_2}}$=$\begin{bmatrix}  
a & b \\ c & d
\end{bmatrix}$=$\gamma$ $\begin{bmatrix}  
1 & S \\ T & 0
\end{bmatrix}$+$\begin{bmatrix}  
d & d \\ d & d
\end{bmatrix}$\,;\,$\gamma>0,\,\&\, a,b,c,d \in \mathbb{R}$\,.\hspace{0.5cm}
\end{tabular}
\label{eq:PayOff_pi}
\end{eqnarray}
More compactly, ${\sf{\Pi}}_2=\gamma{\sf{A}}+d{\sf{1}}$. For the sake of convenience, we have worked with only non-negative values of  $d$. A parallel study for negative $d$ can, in principle, always be  done.

There is a very simple physical significance of $\gamma$ that we discuss now. The two forms of discrete replicator equation can be written as~\cite{taylor1978evolutionary}:
\begin{equation}
\frac{\Delta x}{x}=W({x})\left((\textsf{A}\textbf{x})_1-\textbf{x}^T \textsf{A}\textbf{x}\right),
\label{eq:W}
\end{equation}
where $\Delta x=x'-x$ and `weight function' $W(\textbf{x})> 0$ for $x\in [0,1]$. For type-I equation, $W(x)=\gamma$ and for type-II equation, $W(x)={1}/({\textbf{x}^T \textsf{A}\textbf{x}+d/\gamma})$. Eq.~(\ref{eq:W}) can be interpreted as follows: the L.H.S. denotes the fractional change in the frequency of state $x_1$ and $\left((\textsf{A}\textbf{x})_1-\textbf{x}^T \textsf{A}\textbf{x}\right)$ in the R.H.S. denotes the deviation of fitness of state $x_1$ from the average fitness of the population. The weight function $W(x)$ associates a weight to this deviation, meaning it measures how strongly this deviation affects the fractional change in the frequency of the state. In this context, it is apt to say that $\gamma$ acts as \emph{weight of fitness deviation} and will henceforth be termed as such. To understand the essence of this terminology, we may note that the weight functions are monotonically increasing functions of $\gamma$ and hence so are the corresponding fractional changes in the frequency of the states. Eq.~(\ref{eq:W}) reduces to Eq.~(\ref{eq:type-I}) and Eq.~(\ref{eq:type-II}) for $\gamma=1$ and $\gamma=\infty$ (for finite $d$) when used in $W(x)=\gamma$ and $W(x)={1}/({\textbf{x}^T \textsf{A}\textbf{x}+d/\gamma})$ respectively.

\textcolor{black}{For the sake of argument, consider one dimensional continuous replicator equation,
\begin{eqnarray}
\frac{dx}{dt}=x\left[ \left( {\sf{A}} \textbf{x} \right)_{i}-\textbf{x}^{T}{\sf{A}} \textbf{x} \right],\label{eq:cre}
\end{eqnarray}
and the corresponding type-I replicator map as its trivial discretization. Consequently, $W(x)$ or $\gamma$ in the type-I can be interpreted as the fixed time interval over which total payoffs accumulates before the next frequency-update happens. However, for the type-I map describing non-overlapping generations, such interpretation of $\gamma$ is not relevant; it is a free parameter that essentially highlights the fact that the dynamics of replicator depends on the absolute values (and thus on the relative ordering) of the payoffs rather than merely on their relative ordering. In continuous replicator equation, $\gamma$ can be absorbed in the time derivatives of the frequencies by rescaling time and hence is of no practical importance but in the discrete versions this rescaling is not possible and the evolutionary outcomes depend crucially on the weight of fitness deviation as we witness in the following subsections. That said, one can interpret (two times) $\gamma$ as the inverse temperature (interpreted as intensity of selection) appearing in the Fermi--Dirac distribution function giving the probability that a pure strategy is replaced by the other in a pairwise comparison process~\cite{traulsen2006pre}. This process reduces to Eq.~(\ref{eq:cre}) in the limit of infinite population and weak selection.}

In what follows, we discuss, among other results, how the dynamics of the replicator maps can be modified to bring about strict physical chaotic solutions which may not be otherwise realised in the corresponding cardinally equivalent games.
\subsection{Type-I replicator equation}
Type-I replicator equation, $x'_i=f_i(x_1,x_2)$, for the payoff matrix ${\sf{\Pi_{2}}}$ can be rewritten with:
\begin{equation}
{f}_{1}(x_{1},x_{2})=x_{1}+ \gamma x_{1}\left( \left( {\textsf{A}} \textbf{x} \right)_{1}- \textbf{x}^{T}{\textsf{A}} \textbf{x} \right), \,{f_{2}}=1-{f_{1}}\,.
\label{eq:type-IP2}
\end{equation}
As before, it suffices to work with $x=x_1$ only and so we write
\begin{equation}
\bar{f}_{I}(x)=\gamma(S+T-1)x^3+\gamma (1-2S-T)x^2+(1+\gamma S)x\,,
\label{eq:type-IP2Expand}
\end{equation}
such that $x'=\bar{f}_{I}(x)$. 

The transformations $S \to S/\gamma,\,(T-1) \to (T-1)/ \gamma$ makes $\bar{f}_I\left(x\right)={f}_I\left(x\right)\,$. Therefore, as illustrated in Fig.~\ref{fig_3}, changing $\gamma$ shifts and scales the strict physical region in $S$-$T$ parameter space keeping the overall picture topologically intact. However, more importantly, any change in $\gamma$ brings about change in the domains of $S$-$T$ parameter space for each type of attractors for the replicator dynamics. Thus, strict physical chaos may occur in a Chicken game when the weight of fitness deviation is sufficiently large---something absent if only ${\sf A}$ ($\gamma=1\,,d=0$) is used.
\subsection{Type-II replicator equation}
Type-II replicator equation for payoff matrix ${\sf{\Pi}}_2$ is given as: 
\begin{equation}
f_{1}(\textbf{x})=x_{1}\frac{{\gamma (\textsf{A}}\textbf{x})_1+d}{\gamma \textbf{x}^T{\textsf{A}}\textbf{x}+d}\,,\, f_{2}(\textbf{x})=1-f_{1}(\textbf{x})\,.
\label{eq:type-IIP2}
\end{equation}
Type-II replicator equation gives strict physical solution if $a,b,c,d\ge 0$ or $a,b,c,d\le 0$ (see Eq.~\ref{eq:PayOff_pi}) simply because the positive denominator of non-negative $f_{1}$ is the positive numerator plus a positive part, making $f_{1}\in[0,1]\,\forall\,x_1,x_2\in[0,1]$. As usual, we write type-II equation conveniently as follows:
\begin{equation}
x'=\bar{f}_{II}(x)=x\frac{x(1-S)+S +d/\gamma }{x^2+(S+T)x(1-x)+d/\gamma }\,.
\end{equation}
It should be observed that now the strict physical region is given by $S\ge -d/\gamma$ and $T\ge -d/\gamma$ and with decrease in the value of $\gamma$ the extent of the region increases. The fixed points are $x=0$, $x=1$, and $x=x_{(m)}$ whose physical nature and stabilities depend on which ordinal class of games the replicator equation is based on. Initial conditions in the appropriate basin of attraction converges to a evolutionary stable state (one of the fixed points) in the strict physical region.  We note that in the strict physical region $\bar{f}_{II}$ is a monotonic nondecreasing function of $x$ in simplex. Thus,   arguments analogous to what has been given in Section~\ref{sec:3.B}, periodic orbits or chaotic orbits are absent in type-II replicator equation for any payoff matrix ${\sf \Pi_2}$ chosen from the strict physical region. \textcolor{black}{It is further worth noticing that from equation~(\ref{eq:type-IIP2}) we can give a physical meaning to $d/\gamma$: it is the fitness in the absence of any interaction.}
\subsection{Summary and comparison}
Games cardinally equivalent to any of the twelve classes of games discussed in this paper can have potentially different replicator dynamics. Specifically, it is seen that depending on the weight of fitness deviation ($\gamma$), strictly physical chaotic dynamics can be introduced in some of the classes of games that do not show such behaviour for $\gamma=1$ in type-I replicator equation. Likewise, chaos may be eliminated from a particular game with specific ${\sf A}$ by changing the weight of fitness deviation. Contrary to type-I, the implications of cardinal transformation of ${\sf A}$ does not lead to any new intriguing results in type-II: there still is no strictly chaos or periodic orbits in strict physical region. However, as the weight of fitness deviation ($\gamma$) is decreased then the strict physical region expands to engulf even those games that were earlier not inside the strict physical region. In fact, in the limit $\gamma\rightarrow0$, the strict physical region tends to $\mathbb{R}^2$. Similar limiting behaviour is also witnessed for the case of type-I replicator equation as $\gamma$ is decreased. Thus, taking $\gamma\rightarrow0$ pushes the chaotic region (realised at the tip of the leaf-like strict physical region) towards infinity making detection of chaos rather difficult unless one chooses very large values for $S$ and $T$. In some sense, this means that the discrete replicator equation approaches a continuous version of the replicator equation which anyway, being a one-dimensional autonomous flow, is not supposed to possess chaotic solutions. In the similar vein, both the discrete replicator equations approach the continuous version as the weight of fitness deviation approaches zero and thus it makes sense that the strict physical regions in both the cases becomes $\mathbb{R}^2$  in the limit; of course, $\mathbb{R}^2$ is the strict physical region for the continuous replicator equation.
\textcolor{black}{\section{Is Type-II a better model than Type-I?}
The answer to this question is an emphatic `no'. Although it may sound a bit surprising when seen against the backdrop of the fact that type-II replicator map is much more commonly used in the literature, we argue in following three subsections why type-I is as good or as bad a model as type-II.
\subsection{No control over payoff matrix}
The games with the Prisoner's dilemma's pay-off matrix structure have been observed in the evolution of competitive interactions in RNA viruses~\cite{turner1999nature}, cancer cells~\cite{kareva2011plosone, west2016cspo}, alliance between firms in retail market~\cite{binner2003games}, Hobbes's state of nature~\cite{alexandra1992philosophy}, and in many other innumerable topics of biology, economics, and psychology. Similar omnipresence is that of Stag-hunt game: e.g., emergence of trust-behaviour in population~\cite{fang2002group}, evolution of social structure~\cite{skyrams2004cambridge}, evolutionary dynamics of collective actions in animal world~\cite{pacheco2008proceeding}. We need to appreciate the fact that as scientists we do not have control on the payoff matrices as they would appear in natural scenario; all we can do is to model a given natural scenario having an inherent payoff matrix. The models need to be sensible in not yielding unphysical solutions. It thus is crystal clear that type-II replicator map is at disadvantage when employed to model the discrete replicator dynamics involving payoff matrices of ubiquitously found Prisoners' dilemma and Stag-hunt games simply because the map then gives unphysical solutions. Better model in such cases is obviously type-I map. A relook at Fig.~\ref{fig_1} and similar considerations suggest there are games where however type-I map is a bad model but type-II is sensible one. Hence, we conclude that having no control on the payoff matrices means that both type-I and type-II models can turn out to be unphysical and useless models in certain natural scenarios.} 

\textcolor{black}{Of course, one may argue that in the context of evolutionary biology, the fitness of a trait should be interpreted as the expected number of offspring with that very trait~\cite{Weissing1991evolutionary}. This means that the payoff matrix elements must all be non-negative (in our case $S,T\ge0$) so that not only the fitnesses are non-negative but also the dynamics gives physical phase trajectories. However, as emphasized earlier, the replicator equations have far wider applicability than just mathematizing Darwinian evolution; the replicator equation is used in economics, social sciences, etc., where some elements of the payoff matrix may be taken to be negative rendering type-I map naturally useful in modeling the corresponding dynamics.}

\textcolor{black}{Interestingly, the type-I equation may be seen as a simplified version of the type-II equation as follows: The weight function, $W(x)$, of the later is frequency dependent and decreases with the increase in the average payoff of the population. This means that for a given difference between the expected payoff of a pure strategy and the average fitness of the population, the relative change in frequency is comparatively less for the case in which the average fitness of the population is more. Critically speaking, this appears to be an artificial condition added to the basic tenet of Darwinism. Interestingly, from the expressions of $W(x)$ for the two maps, one can easily conclude that if we simplify the type-II model by imposing the condition that the relative change in the frequency be independent of the average fitness of population (for a given excess of fitness of a strategy over the average fitness), we essentially arrive at the type-I model. It must be noted, probably not surprisingly, that the aforementioned independence is inherent in the continuous replicator equation.}
\textcolor{black}{\subsection{Theorems relating NE to attractors}
In population biology, folk theorems---relating evolutionary outcome of a system to the NE behaviour---have facilitated paradigm shift towards strategic reasoning~\cite{cressman2014replicator}, practically elevating the players to the status of rational decision makers. However, a parallel well-known fact is that emergence of chaos (either in continuous or discrete models) means that rationality is probably unrealistic approximation~\cite{sato2002chaos}.  By now we have seen that in the first quadrant in $S$-$T$ space while type-II map doesn't show chaos or oscillatory dynamics, type-I map does; both however have fixed point solutions. This doesn't at all mean that type-II is a better model. In fact, it just means that type-II is dynamically less rich. Of course, in order to claim that type-I map in on similar footing as type-II map, we must relate its (fixed point) solutions to the NE and following theorems (see Appendix~C for proofs) accomplish exactly that:
\begin{enumerate}
\item If $\hat{\bf x} \in \Sigma^n$ is a NE of the population game described by the payoff matrix $\sf{\Pi}$, then $\hat{\bf x}$ is a fixed point of type-I replicator map.
\item If $\hat{\bf x}$ is $\omega$-limit of an orbit ${\bf x}(t_n)\in \textrm{int} \Sigma^n$, then $\hat{\bf x}$ is a NE.
\item If a fixed point, $\hat{\bf x}$, of type-I replicator map is Lyapunov stable then it is NE.
\end{enumerate}
Thus, even in this respect the type-I model is as good as the type-II model.}
\textcolor{black}{\subsection{Equilibrium concept in game theory and $m$-period orbits}
We know that 1-period stable orbits, i.e., fixed point attractors, in discrete replicator dynamics are evolutionary stable states and hence, by definition, Nash equilibria. However, since the type-1 map is capable of showing $m$-period attractors, it is very important to find their connection with game-theoretic equilibrium concepts. To this end, we note that in case ${\bf x}^{(1)}$ (this choice of notation is for later convenience) is a mixed NE, following is valid:
\begin{eqnarray}
({\sf\Pi}{\bf x}^{(1)})_i={{\bf x}^{(1)}}^T{\sf \Pi}{\bf x}^{(1)},\,\forall i\in\{1,\cdots,n\},\label{eq:ar1}
\end{eqnarray}
which also means that ${\bf x}^{(1)}$ is a fixed point or 1-period point. Now, multiplying Eq.~(\ref{eq:ar1}) with $x_i^{(1)}$ we get the equivalent condition for mixed NE:
\begin{equation}
x_i^{(1)}({\sf\Pi}{\bf x}^{(1)})_i=x_i^{(1)}[{{\bf x}^{(1)}}^T{\sf \Pi}{\bf x}^{(1)}].\label{eq:ar2}
\end{equation}
The physical meaning of this equation is that the normalized total expected payoff of the entire set of $i$th phenotypic individuals, making fraction $x_i$ of the population, is equal to the normalized average total payoff of any randomly chosen fraction ($x_{\rm (any)}$) of individuals (not necessarily of a common phenotype) such that $x_{\rm (any)}=x_i$ for any allowed $i$. We have merely recast the interpretation of the NE with the benefit that this interpretation allows us to logically extend the concept of the equilibrium in the following way: Suppose there exists an $m$-period orbit---$\{{\bf x}^{(1)},\,\textbf{x}^{(2)},\cdots,\,\textbf{x}^{(k)},\cdots,\,\textbf{x}^{(m)}\}$, i.e.,
\begin{equation}
x^{(k+1)}_i=x^{(k)}_i+x^{(k)}_i\left[({\sf\Pi}\textbf{x}^{(k)})_i-{\textbf{x}^{(k)}}^T {\sf\Pi}\textbf{x}^{(k)}\right],
\label{eq:ar3}
\end{equation}
for all $k\in\{1,2,\cdots,m\}$ and ${\bf x}^{(m+1)}\equiv {\bf x}^{(1)}$.}

\textcolor{black}{Now, consider the following generalization of NE or Eq.~({\ref{eq:ar2}}):
\begin{equation}
\sum_{k=1}^m x_i^{(k)}({\sf \Pi}{\bf x}^{(k)})_i=\sum_{k=1}^m x_i^{(k)}[{{\bf x}^{(k)}}^T{\sf \Pi}{\bf x}^{(k)}],\label{eq:ar4}
\end{equation}
and the obvious extension of the aforementioned physical meaning, now over $m$ generations, that applies to Eq.~(\ref{eq:ar4}). By adding all the $m$ equations implied by Eq.~(\ref{eq:ar3}), we see that Eq.~(\ref{eq:ar4}) is also the necessary condition for the existence of the $m$-period orbit. Thus, we have connected the game-theoretic concept of the extended NE with the evolutionary outcome---$m$-period orbits of type-I replicator map---of the corresponding dynamical system.}
\section{Discussion and Conclusion}
Replicator dynamics is best realised in an infinitely large well-mixed unstructured population. In this paper, we have theoretically investigated and contrasted the dynamics of the two types of most commonly used discrete replicator equations---type-I and type-II. In principle, this study could be done with any number of players and strategies. Unlike for the case of continuous replicator equations, the discrete replicator equations are capable of showing rich dynamics including chaos even for two-player, two-strategy scenario. Hence, in this paper we have focussed on this simplest possible nontrivial replicator equations in order to understand the interplay of chaos and the underlying simple games that governs the evolutionary dynamics dealing with the interaction and subsequent differential reproduction of individuals in a population. As emphasised again and again in this paper, it is more tricky to interpret the solutions of the discrete replicator equations because for some parameter values, the variables (normalised frequencies of traits) takes values outside the interval $[0,1]$ which is physically meaningless. This motivated us to analytically find the \emph{sufficient conditions} for the existence of the strict physical regions in the parameter space such that any trajectory in the state space of the corresponding replicator equation always remains meaningfully bounded: the frequencies of the traits can neither have negative values nor values greater that unity. In the process we find that among the twelve classes of ordinally equivalent games, on which the replicator equations are based, only games \textbf{ii}-\textbf{iv} (Chicken games, Leader games, and games of Battle of sexes) and games \textbf{vi}-\textbf{viii} (which includes Harmony games) lie in the strict physical region of type-II replicator equation. For type-I equation two more games---Prisoner's dilemma (\textbf{i}) and Stag-hunt games (\textbf{v})---may show up in the corresponding strict physical region depending on the parameter values. We also note that while physical chaos is not witnessed in the dynamics of type-II equation in the strict physical region, it does show up in the dynamics of type-I equation. However, in $S$-$T$ parameter space the region corresponding to chaos is highly localised near the tip of the leaf-like structure (see Fig.~\ref{fig_2}a). This means that type-I replicator equation based on payoff matrices of most games, except Leader game and Battle of sexes, do not show chaos for the parameter values in the strict physical region.

Interestingly, on invoking the concept of the weight of fitness deviation, we discover the following important facts: Firstly, the extent of the strict physical region increases with the decrease in the weight of fitness deviation; and secondly, games (other than Leader game and Battle of sexes) such as Chicken game may start showing (strict) physical chaotic solutions in the corresponding replicator dynamics as the weight of fitness deviation increases. Thus, we note that it is $\gamma$ which governs the existence of physical chaos for a given value of $S$ and $T$. Similarly, for the case of type-II equation, decreasing $\gamma$ sufficiently makes any of the twelve classes of games come into the strict physical region for the equation. But, of course, no chaotic behaviour is still observed. As a byproduct of our study we have also proven that the chaos reported~\cite{vilone2011chaos} in type-I replicator with Chicken game's payoff matrix cannot occur in the strict physical region. 

Replicator equations are traditionally used to model the evolutionary game dynamics in infinite population so that stochastic effects can be ignored. It is encouraging that though more realistic stochastic approaches includes the effects of finite population size effects, the deterministic dynamics has been successfully invoked many times to gain basic understanding of the system. There is another simplifying assumption: each individual interacts with every other individual in the population. On relaxing this assumption, meaning on introducing different numbers of interactions for different individuals, we expect the payoff matrix elements to be stochastic in nature. This is another way of introducing stochasticity in the game dynamics. In contrast to the randomness introduced in the dynamics of the variable due to stochasticity in various ways, simple deterministic discrete replicator equations introduce unpredictability in the dynamics owing to chaos that in turn owes its origin to the nonlinear nature of the replicator equations. Nash equilibria and evolutionary stable states are no longer decisive when the system is chaotic. In fact, replicator equations are also used to model reinforcement learning where it may be argued~\cite{sato2002chaos} that chaos is a necessary condition for intelligent adaptive players to fail to converge to a NE. In passing, it may be mentioned that using local replicator equation~\cite{hilbe2011replicator} one may connect evolutionary models for infinite and finite populations when the population itself is infinite but interactions and reproduction occur in random groups of finite size. Intriguingly, the local replicator dynamics is effectively the traditional replicator system with a slightly modified payoff matrix.

Replicator equation does not take mutation into account and hence our investigation may be extended for replicator-mutator equation in order to include the effects of mutation on the extent of the strict physical region and the chaotic solutions therein. Such a research with the discrete versions of replicator-mutator equation will ascertain the role of chaos in the deterministic evolution of the universal grammar that specifies the mechanism of language acquisition~\cite{nowak2001grammar, komarova2001evolutionary}. Also, investigation done in this paper can be extended for two player asymmetric games which in the continuous case are known to occasionally exhibit Hamiltonian chaos~\cite{sato2002chaos}. 

Before we end, it should be realised that the replicator equations are mere models of evolutionary dynamics (and many other analogous systems). Whether they are good models can only be justified through compatible observations and experiments. If the system under consideration is showing irregular behaviour (like non-convergence to NE, bounded but ever-wandering phase trajectories, etc.) and it is wished to model it using deterministic models, then one should choose the model that is capable of showing chaos---specifically, type-I equation and not type-II equation. In the continuous version, a necessary condition for chaotic solutions to appear is that the square payoff matrix \textsf{A} should be at least four dimensional. Hence, it is more cumbersome to model chaotic dynamics by using continuous replicator equations than by using the discrete ones for which even one-dimensional state space is enough. However, in spite of this advantage of discrete equations, it also has to be justified on physical grounds why and when one should use a discrete equation and not the continuous version of it. An answer to this question might be given on case to case basis. While one obvious case for using type-II replicator equation is when the generations of a population are approximately non-overlapping~\cite{cole1954population,godfrey1989discrete}, another one could be when one may want to model a discretely sampled data of a population for which the relevant variables are not known {\em a priori}.

\section*{ Acknowledgments} The authors are thankful to Aditya Tandon for helpful discussions. S.C.~acknowledges the financial support through the INSPIRE faculty award (DST/INSPIRE/04/2013/000365) conferred by the Indian National Science Academy (INSA) and the Department of Science and Technology (DST), India.

\appendix
\section{\textcolor{black}{The leftover games}}
\label{Ap:1}
\textcolor{black}{We may note that the form of $\textsf{A}$, although has the advantage that one needs to worry about only two independent parameter, viz., {$S$} and $T$, it definitely doesn't take into account the games with payoff matrix of the form}
\begin{equation}
\textcolor{black}{
{\sf{\Pi}}={\sf{B}}=:
\begin{bmatrix}
    0      & \tilde{S} \\
    \tilde{T}      & 0 \\
\end{bmatrix};\quad \tilde{S},\tilde{T} \in\mathbb{R}.}
\end{equation}
\textcolor{black}{The corresponding type-1 replicator equation is}
\begin{equation}
\textcolor{black}{\tilde{f}_{I}(x)=(\tilde{S}+\tilde{T} )x^3+[ -\tilde{T}-2\tilde{S}]x^2+[1+\tilde{S}]x\,.}
\end{equation}
\textcolor{black}{It is interesting to note that the transformation:}
\begin{equation}
\textcolor{black}{\tilde{S}\rightarrow S,\quad \tilde{T} \rightarrow T-1,}
\end{equation}
\textcolor{black}{gets us the replicator equation back with ${\sf \Pi}={\sf A}$ [see Eq.~(\ref{eq:type-IAExpand})]. Thus, all the results pertaining to the replicator map with payoff matrix $\textsf{A}$ can be trivially extended to the case of the replicator map with payoff matrix $\textsf{B}$ by mere replacement of $S$ and $T$ by $\tilde{S}$ and $\tilde{T}+1$ respectively.}

\textcolor{black}{Again, with ${\sf{B}}$, the type-II replicator map is,}
\begin{equation}
\textcolor{black}{\tilde{f}_{II}(x)=\frac{x(1-x)\tilde{S}}{x(1-x)(\tilde{S}+\tilde{T})}.}
\end{equation}
\textcolor{black}{From the form of this map, it is clear that the region given by $(\tilde{S} \ge 0; \tilde{T} \ge 0)$ is the strict physical region because the numerator of the righthand side is never greater than the denominator of the righthand side.}
\textcolor{black}{\section{Stable fixed point of type-I replicator equation is evolutionary stable state}
\label{Ap:2}
We have for two-player, two-strategy type-I discrete replicator equation:
\begin{equation}
x'= x+x\left[(\textsf{A}\textbf{x})_1-\textbf{x}^T \textsf{A}\textbf{x}\right].
\end{equation}}

\textcolor{black}{\emph{To prove}: If $\hat{x}$ is a stable fixed point of above equation, then the corresponding state $\hat{\textbf{x}}=(\hat{x},1-\hat{x})$ is evolutionary stable state, i.e., $\exists\,\mathcal{B}_{\hat{\textbf{x}}}\subset\Sigma^2$ such that $\forall \textbf{y} \in \mathcal{B}_{\hat{\textbf{x}}}\backslash\{{\hat{\textbf{x}}}\}$, $\hat{\textbf{x}}^{T}\textsf{A}\textbf{y}>\textbf{y}^T\textsf{A}\textbf{y}$.}

\textcolor{black}{\emph{Proof:} If $\hat{{x}}$ is a stable fixed point of map then there exists a neighbourhood, $\mathcal{N}_{\hat{x}}$ of $\hat{x}$ in $(0,1)$  such that $\forall\,y\in\mathcal{N}_{\hat{x}}\backslash\{\hat{x}\}$ we have:
\begin{eqnarray}
&&\frac{||{y}'-\hat{{x}}||}{||{y}-\hat{{x}}||}<1\,,\\
{\rm or,}\,\,&&\frac{||y-\hat{x}+y\left[(\textsf{A}\textbf{y})_1-\textbf{y}^{T}\textsf{A}\textbf{y}\right]||}{||y-\hat{x}||}<1\,,\\
{\rm or,}\,\,&&\frac{||y-\hat{x}+y(1-y)\left[(\textsf{A} \textbf{y})_1-(\textsf{A}\textbf{y})_2\right]||}{||y-\hat{x}||}<1\,.\qquad
\label{eq:stabilitycondition}
\end{eqnarray}
Here, $||\cdots||$ stands for an appropriate norm which we can conveniently take as the Euclidean norm. Inequality~(\ref{eq:stabilitycondition}) implies that $y-\hat{x}$ and $(\textsf{A}\textbf{y})_1-(\textsf{A}\textbf{y})_2$ must have opposite signs. Therefore if $\hat{x}$ is stable then there exists a neighbourhood in $\Sigma^2$, viz., $\mathcal{B}_{\hat{\bf x}}=\mathcal{N}_{\hat{x}}\times(0,1)\backslash\mathcal{N}_{\hat{x}}$ such that 
\begin{eqnarray}
&&(y-\hat{x})\left[(\textsf{A}\textbf{y})_1-(\textsf{A}\textbf{y})_2\right]<0 \,\forall \,(y,1-y) \in B_{\hat{\bf x}},\qquad\\
\Leftrightarrow\quad &&\hat{\textbf{x}}^{T}\textsf{A}\textbf{y}>\textbf{y}^T\textsf{A}\textbf{y},
\end{eqnarray}
i.e., $\hat{\textbf{x}}$ is evolutionary stable state (and hence also NE).}
\textcolor{black}{\section{Proofs of theorems relating NE to fixed points}}
\label{Ap:3}
\textcolor{black}{The following proofs for type-I equation closely follow its continuous counterpart.}
\subsection{\color{black}{If $\hat{\bf x} \in \Sigma^n$ is a NE of the population game described by the payoff matrix $\sf{\Pi}$, then $\hat{\bf x}$ is a fixed point of type-I replicator map.}}
\textcolor{black}{If $\hat{\bf x}$ is a NE, then $\forall\,{\bf x}\in\Sigma^n$, ${\hat{{\bf x}}}^T\sf{\Pi}\hat{{\bf x}}\ge{\bf x}^T\sf{\Pi}\hat{{\bf x}}$. We denote the corners of the simplex with standard unit vectors ${\bf e}_i$ each of which corresponds to a pure strategy. Choosing ${\bf e}_i$ as ${\bf x}$, we arrive at $({\sf\Pi}\hat{\textbf{x}})_i=\textrm{constant}\,\forall\,i$ such that $x_i>0$. This is nothing but the condition for $\hat{\bf x}$ to be the fixed point in the interior of the simplex.}

\textcolor{black}{For the case when $\hat{\bf x}$ is a strict NE, $\hat{\bf x}={\bf{e}}_i$ for  $i=k\, (\textrm{say})$, since strict Nash strategy must be pure. This means, by definition, ${\hat{{\bf x}}}^T\sf{\Pi}\hat{{\bf x}}=(\sf{\Pi}\hat{{\bf x}})_k$ which implies that ${\bf{e}}_k$ is the fixed point as well.}
\subsection{\color{black}{If $\hat{\bf x}$ is $\omega$-limit of an orbit ${\bf x}(t_n)\in \textrm{int} \Sigma^n$, then $\hat{\bf x}$ is a NE.}}
\textcolor{black}{Assume that $\hat{\bf x}$ is an $\omega$-limit of an orbit $\textbf{x}(t_n)$ in $\textrm{int} \Sigma^n$ but is not a NE. Then $\exists\epsilon > 0$ and an $i \in \{1,\cdots,n\}$ such that, $(\sf{\Pi}{\textbf{x}})_i-{\textbf{x}}^T\sf{\Pi}{\textbf{x}}>\epsilon$. (This means that ${\bf x}$ in the neighbourhood of $\hat{\bf x}$ and on the orbit, is not a NE; had it been a NE then it must have been a fixed point thereby restricting the orbit from further approach towards the $\omega$-limit, $\hat{\bf x}$.) This implies ${\Delta {x}_i}/{{x}_i}>\epsilon$, meaning that for sufficiently large interval of time an orbit ${\bf x}(t_n)$ diverges away from $\hat{\textbf{x}}$. This is a contradiction since $\hat{\textbf{x}}$ is the $\omega$-limit of an orbit ${\bf x}(t_n)\in \textrm{int} \Sigma^n$. Thus, $\hat{\textbf{x}}$ must be a NE.}
\subsection{\color{black}{If a fixed point, $\hat{\bf x}$, of type-I replicator map is Lyapunov stable then it is NE.}}
\textcolor{black}{If we assume that $\hat{\bf x}$ is Lyapunov stable but is not a NE, then $\exists\,{\rm an}\, i \in \{1,\cdots,n\}$ such that $({\sf{\Pi}}{\hat{\textbf{x}}})_i-{\hat{\textbf{x}}}^T{\sf{\Pi}}{\hat{\textbf{x}}}>0$. Therefore, in the sufficiently small neighbourbood of $\hat{\bf x}$, one can write for any ${\bf x}$: $\exists\epsilon > 0$ and an $i \in \{1,\cdots,n\}$ such that $({\sf{\Pi}}{\textbf{x}})_i-{\textbf{x}}^T{\sf{\Pi}}{\textbf{x}}>\epsilon$. This implies ${\Delta {x}_i}/{{x}_i}>\epsilon$ implying that ${x}_i$ exponentially diverges with time in the neighbourhood. Thus, $\hat{\textbf{x}}$ is not Lyapunov stable which is a contradiction. Hence, $\hat{\textbf{x}}$ is a NE.}
\bibliography{Pandit_etal_manuscript}
 \end{document}